\documentclass[12pt]{article}
\usepackage{graphics,epsfig}
\usepackage[english]{babel}
\newcommand{\cinst}[2]{$^{\mathrm{#1}}$~#2\par}
\newcommand{\crefi}[1]{$^{\mathrm{#1}}$}

\begin{document}


\begingroup

\raisebox{0.5cm}[0cm][0cm] {
\begin{tabular*}{\hsize}{@{\hspace*{5mm}}ll@{\extracolsep{\fill}}r@{}}
\begin{minipage}[t]{3cm}
\vglue.5cm
\end{minipage}
&
\begin{minipage}[t]{7cm}
\vglue.5cm
\end{minipage}
&

\end{tabular*}
}

\begin{center}

{\Large{\bf A study of the nuclear medium influence\\
 on neutral strange particle production \\
\vspace{0.2cm} in deep inelastic neutrino scattering }}
\end{center}

\vspace{1.cm}

\begin{center}
{\large SKAT Collaboration}

 N.M.~Agababyan\crefi{1}, V.V.~Ammosov\crefi{2},
 M.~Atayan\crefi{3},\\
 N.~Grigoryan\crefi{3}, H.~Gulkanyan\crefi{3},
 A.A.~Ivanilov\crefi{2},\\ Zh.~Karamyan\crefi{3},
V.A.~Korotkov\crefi{2}

\setlength{\parskip}{0mm}
\small

\vspace{1.cm} \cinst{1}{Joint Institute for Nuclear Research,
Dubna, Russia} \cinst{2}{Institute for High Energy Physics,
Protvino, Russia} \cinst{3}{Yerevan Physics Institute, Armenia}
\end{center}
\vspace{100mm}

{\centerline{\bf YEREVAN  2004}}

\newpage
\begin{abstract}

\noindent The influence of nuclear effects on the production of
neutral strange particles ($V^{\circ}$) is investigated using the
data obtained with SKAT propane-freon bubble chamber irradiated in
the neutrino beam (with $E_{\nu}$ = 3-30 GeV) at Serpukhov
accelerator. The measured mean multiplicity of $V^{\circ}$
particles in nuclear interactions, ${<n_{V^{\circ}}>}_A$ =
0.096$\pm$0.011, is found to exceed significantly that in
'quasinucleon' interactions,
 ${<n_{V^{\circ}}>}_N$ = 0.059$\pm$0.012. The ratio of 
${\rho}_{V^{\circ}}$
 =${<n_{V^{\circ}}>}_A/{<n_{V^{\circ}}>}_N$ =1.61$\pm$0.23 is 
 larger than that for $\pi^{-}$ mesons, ${\rho}_{\pi^-}$ =
 1.10$\pm$0.03. It is shown that a dominant part of the
 multiplicity gain of $V^{\circ}$ particles can be explained by
 intranuclear interactions of produced pions.
\end{abstract}


\newpage
\section{Introduction}

The experimental study of inclusive spectra of charged hadrons
produced in the lepton-nucleus deep inelastic scattering (DIS)
exhibits a significant influence of the nuclear medium on the
particle yield in various phase space regions (see
\cite{ref1,ref2} and references therein). The secondary
intranuclear interactions results in a depletion of the fast
hadron yield (in the current quark fragmentation region) and, to a
more extent, an enhancement in the target fragmentation region.
The magnitude of these effects and its dependence on the energy
$\nu$ transferred to the hadronic system or on its invariant mass
$W$ are found \cite{ref1,ref2,ref3} to be in qualitative agreement
with prediction of models \cite{ref4,ref5} considering the
space-time evolution of the quark string fragmentation and the
hadron formation in the nuclear medium.\\ The magnitude of the
nuclear effects on the particle yield can differ for various types
of hadrons, for example, for charged ones (predominantly pions)
and for the neutral strange particles ($V^{\circ}$). Indeed, the
yield of the latter in 'elementary' $l N$ interactions is more
than one order of magnitude lower than for the former (c.f., for
example, the data \cite{ref6,ref7,ref8,ref9} on $\nu N$
interactions). Hence, the secondary intranuclear reactions $\pi N
\rightarrow K^{\circ}X$ and $\pi N \rightarrow{\Lambda}^{\circ}X$,
though their relatively small inclusive cross sections, can cause
a large gain of the $V^{\circ}$ yield in the target fragmentation
region. On the other hand, the depletion of the $V^{\circ}$ yield
in the current quark fragmentation region is expected to be weaker
due to the smaller absorption cross section of strange particles
as compared to that for pions.\\ Hitherto these expectations for
the leptoproduction of strange particles on nuclei are not checked
experimentally. This work is devoted to study of nuclear effects
on the neutral strange particle production in the neutrino-nucleus
DIS at intermediate energies,  $\nu$ = 2-15 GeV and $W$ = 2-5 GeV.
In Section 2, the experimental procedure is briefly described. The
experimental results are presented and discussed in Section 3. The
results are summarized in Section 4.

\section{Experimental procedure}

The experiment was performed with SKAT bubble chamber
\cite{ref10}, exposed to a wideband neutrino beam obtained with a
70 GeV primary protons from the Serpukhov accelerator. The chamber
was filled with a propan-freon mixture containing 87 vol\% propane
($C_3H_8$) and 13 vol\% freon ($CF_3Br$) with the percentage of
nuclei H:C:F:Br = 67.9:26.8:4.0:1.3 \%. A 20 kG uniform magnetic
field was provided within the operating chamber volume. The
selection criteria of the properly reconstructed charged current
interactions and the procedure of the reconstruction of the
neutrino energy $E_{\nu}$ can be found in our previous
publications (\cite{ref1} and references therein). Each event was
given a weight (depending on the charged particle multiplicity)
which corrects for the fraction of events excluded due to
improperly reconstruction. The events with 3 $<E_{\nu}<$ 30 GeV
were accepted, provided that $W>$ 2 GeV and the transfer momentum
squared $Q^2>$ 1 (GeV$/c)^2$. The number of accepted events was
3101 events (3706 weighted events). The mean values of the
kinematical variables are: $<E_{\nu}>$ = 12.2 GeV, $<Q^2>$ = 3.6
(GeV/$c)^2$, $<W>$ = 3.0 GeV, $<W^2>$ = 9.7 GeV$^2$, and $<{\nu}>$
= 6.6 GeV.\\ The selection criteria for the decay of neutral
strange particle and the procedure of their identification were
similar to those applied in \cite{ref11}. The number of the
accepted $V^{\circ}$ was 80 out of which 32(48) had the biggest
probability to be identified as $K^{\circ}({\Lambda}^{\circ})$.
The corresponding average multiplicities, corrected for the decay
losses, are $<n_{V^{\circ}}>$ = 0.092$\pm$0.010, $<n_{K^{\circ}}>$
= 0.053$\pm$0.009 and $<n_{\Lambda^{\circ}}>$ = 0.039$\pm$0.006.\\
For the further analysis the whole event sample was subdivided,
using several topological and kinematical criteria \cite{ref12},
into three subsamples: the 'cascade' subsample $B_S$ with a sign
of intranuclear secondary interactions, the 'quasiproton' $(B_p)$
and 'quasineutron' $(B_n)$ subsamples for which no sign of
secondary interactions was observed. About 40\% of subsample $B_p$
is contributed by interactions with free hydrogen. Weighting the
'quasiproton' events with a factor of 0.6, one can compose a
'pure' nuclear subsample: $B_A = B_S + B_n + 0.6 B_p$ (with an
effective atomic weight $\bar{A}$ = 28) and a 'quasinucleon'
subsample $B_N = B_n + 0.6 B_p$. In the next Section, the
characteristics of neutral strange particles in subsamples $B_A$
and $B_N$ will be compared to infer an information about the
influence of the nuclear medium on their production.

\section{Experimental results}

Due to the lack of statistics, we will consider below the
combined data on $K^{\circ}$ and $\Lambda$. Table 1 shows the mean
multiplicity $<n_{V^{\circ}}>$ and, for comparison, $<n_{\pi^-}>$
for $\pi^-$ meson \cite{ref2}, as well as their ratio
$R(V^{\circ}/\pi^-)$ = $<n_{V^{\circ}}>/<n_{\pi^-}>$ for
subsamples $B_N$ and $B_A$.

\begin{table}[ht]
\begin{center}
\begin
{tabular}{|l|c c c|}
  \hline
Subsample&$<n_{V^{\circ}}>$&$<n_{\pi^-}>$&$R(V^{\circ}/\pi^-)$ \\
\hline $B_N$&0.059$\pm$0.012&0.813$\pm$0.026&0.072$\pm$0.015 \\
$B_A$&0.096$\pm$0.011&0.897$\pm$0.020&0.106$\pm$0.012 \\ \hline
\end{tabular}

\end{center}
\caption{The mean multiplicities $<n_{V^{\circ}}>$ and
$<n_{\pi^-}>$ and their ratio for subsamples $B_N$ and $B_A$.}
\end{table}

\noindent The quoted value of $<n_{V^{\circ}}>_N$ =
0.059$\pm$0.012 for the 'quasinucleon' subsample is consistent
with the data around $W^2\sim$ 10 GeV$^2$ obtained for neutrino -
nucleon interactions (cf. \cite{ref6} and references therein), 
while that for nuclear interactions, 
$<n_{V^{\circ}}>_A$ = 0.096$\pm$0.011, 
is close to the value 0.102$\pm$0.009 measured in
\cite{ref11} for the case of a heavier target (freon), but at
somewhat lower $<W^2>$ = 7.8 GeV$^2$.  
The relative yields
${R_N}(V^{\circ}/\pi^-)$ and ${R_A}(V^{\circ}/\pi^-)$ in
subsamples $B_N$ and $B_A$ differ by about 1.5 times, i.e. the
production of neutral strange particles is influenced by the
nuclear medium stronger than that for pions.\\ The
multiplicity gain for $\pi^-$ mesons, characterized by the ratio
$\rho_{\pi^-} = <n_{\pi^-}>_A/<n_{\pi^-}>_N$ = 1.10$\pm$0.03, can
be qualitatively explained \cite{ref1} by the secondary
intranuclear interactions of produced pions, taking into account
the finiteness of their formation lenght \cite{ref4}. This ratio
for $V^{\circ}$ is equal to $\rho_{V^{\circ}}$ = 1.62$\pm$0.23.
(Note, that in the evolution of the error of the ratio $\rho =
<n>_A/<n>_N$, the corrrelation between $<n_A>$ and $<n_N>$ was
taken into account). If the gain $\rho_{V^{\circ}}$ is caused by
intranuclear interactions of produced pions, i.e. by the secondary
reactions $\pi N \rightarrow V^{\circ} + X$, then one can expect
that the products of latter will predominantly occupy the backward
hemisphere of the hadronic c.m.s., i.e. the region of $x_F < 0$
($x_F$ being the Feynman variable). This expectation is
qualitatively supported by the data presented in Table 2 for two
regions of $x_F$: $x_F > 0$ and $x_F < 0$. The data indicate that
the nuclear enhancement of the $V^{\circ}$ yield occurs mainly
in the backward hemisphere.

\begin{table}[ht]
\begin{center}
\begin
{tabular}{|l|c c c|}
  \hline
Subsample&$<n_{V^{\circ}}>$&$<n_{\pi^-}>$&$R(V^{\circ}/\pi^-)$ \\
\hline &\multicolumn{3}{|c|}{$x_F > 0$}\\

 $B_N$&0.025$\pm$0.008&0.474$\pm$0.019&0.052$\pm$0.017 \\
$B_A$&0.034$\pm$0.009&0.437$\pm$0.013&0.077$\pm$0.021 \\ \hline

&\multicolumn{3}{|c|}{$x_F < 0$}\\

 $B_N$&0.034$\pm$0.009&0.339$\pm$0.018&0.101$\pm$0.027 \\
$B_A$&0.062$\pm$0.010&0.460$\pm$0.015&0.134$\pm$0.022 \\ \hline

\end{tabular}

\end{center}
\caption{The mean multiplicities $<n_{V^{\circ}}>$ and
$<n_{\pi^-}>$ and their ratio in regions $x_F > 0$ and $x_F < 0$
for subsamples $B_N$ and $B_A$.}
\end{table}

\noindent The ratio $\rho_{V^{\circ}}$ at $x_F < 0$ and $x_F > 0$
is equal, respectively, 
$\rho_{V^{\circ}}(x_F < 0)$ = 1.80$\pm$0.32 and
$\rho_{V^{\circ}}(x_F > 0)$ = 1.37$\pm$0.35.
These values differ from those for 
$\pi^-$ mesons: $\rho_{\pi^-}(x_F < 0)$ = 1.36$\pm$0.05
and $\rho_{\pi^-}(x_F > 0)$ = 0.92$\pm$0.03. Unlike $V^{\circ}$'s,
the $\pi^-$ yield at $x_F > 0$ is depleted, while the enhancement
at $x_F < 0$ is significantly smaller. These depletion and enhancement
effects for $\pi^-$ mesons can be explained by inelastic
interactions of produced pions ($\pi^+, \pi^-, \pi^{\circ}$)
inside the nucleus. The calculations in the framework of a model
considered in \cite{ref1} result in $\rho_{\pi^-}^{th}(x_F < 0)$ =
1.65$\pm$0.14 and $\rho_{\pi^-}^{th}(x_F > 0)$ = 0.98$\pm$0.05,
being in a reasonable agreement with the data.

\noindent Below an attempt is undertaken to estimate the
multiplicity gain, ${\delta}_{V^{\circ}}({\pi}N) =
<n_{V^{\circ}}>_A - ~~<n_{V^{\circ}}>_N$, caused by intranuclear
interactions of 'primary' pions, ${\pi}N \rightarrow V^{\circ} +
X$, resulting in production of secondary $V^{\circ}$ particles.
The mean multiplicity of the latter in inelastic ${\pi}N$
interactions (averaged over protons and neutrons of the target
nuclei), being estimated from the available experimental data
\cite{ref13}, increases from ${\overline{n}}_{V^{\circ}}(p_{\pi})
\approx$ 0.03 (0.06) for $\pi^+(\pi^-)$ induced reactions at the
pion momentum $p_{\pi} \sim 1$ GeV$/c$ up to 0.15 at $p_{\pi}
\sim$ 15 GeV$/c$ (above which the fraction of pions produced in
the ${\nu}N$ DIS is negligible in this experiment). The
probability of the secondary inelastic interactions of pions
averaged over their formation length (taken from \cite{ref4}) and
the nuclei of the propane-freon mixture is estimated to be
$w_{in}(p_{\pi})$ = 0.24 at $p_{\pi} = 1-2$ GeV$/c$ decreasing up
to $w_{in}(p_{\pi})$ = 0.15 at $p_{\pi} \sim 10-15$ GeV$/c$. The
integration of the product $w_{in}(p_{\pi}) \cdot
{\overline{n}}_{V^{\circ}}(p_{\pi})$ over the pion momentum
spectrum from $p_{\pi}$ = 0.9 GeV$/c$ to 15 GeV$/c$ results in
${\delta}_{V^{\circ}}({\pi}^{+}N) = (1.5 \pm 0.23) \cdot 10^{-2}$
and ${\delta}_{V^{\circ}}({\pi}^{-}N) = (0.50 \pm 0.07) \cdot
10^{-2}$ for ${\pi}^+$ and ${\pi}^-$- induced reactions,
respectively, where the quoted errors reflect the uncertainty in
${\overline{n}}_{V^{\circ}}(p_{\pi})$ \cite{ref13}. \\ The
contribution from ${\pi}^{\circ}$- induced reactions,
${\pi}^{\circ} N \rightarrow V^0 + X$, is assumed to be an average
of those for ${\pi}^+$ and ${\pi}^-$ mesons,
${\delta}_{V^{\circ}}({\pi}^{\circ}N) = 0.5
[{\delta}_{V^{\circ}}({\pi}^{+}N) +
{\delta}_{V^{\circ}}({\pi}^{-}N)]$, with an uncertainty $\pm
0.5[{\delta}_{V^{\circ}}({\pi}^{+}N) +
{\delta}_{V^{\circ}}({\pi}^{-}N)]$. The resulting expected gain of
the $V^{\circ}$ multiplicity turns out to be
${\delta}_{V^{\circ}}({\pi} N) = (0.030 \pm 0.007)$ which is in
reasonable agreement with the experimental value
${\delta}_{V^{\circ}}^{exp} = 0.037 \pm 0.010$ (cf. Table 1).\\
Note finally, that the applied model of intranuclear
interactions is rather crude and uses several simplified assumptions
some of which should be pointed out: i) the
second-order effects of more than one inelastic collisions of
pions are neglected; ii) the charge-exchange reactions like $K^+ n
\longleftrightarrow K^{\circ} p$, ${\sum}^+ n \longleftrightarrow
{\Lambda} p$ and so on are not considered; iii) the model does not
incorporate the production of hadronic resonances with a proper
space-time structure of their formation, intranuclear interactions
and decay. Hence, more refined model calculations and
statistically more provided experimental data are needed to
establish whether the nuclear strangeness enhancement observed in
this work could be fully attributed to secondary intranuclear
interactions or other nuclear effects play a non-negligible role.

\section{Summary}

For the first time, the influence of the nuclear medium on the
neutrinoproduction of neutral strange particles is studied at the
energy range $3 < E_\nu < 30$ GeV. The mean multiplicity of $V^0$
particles in nuclear and 'quasinucleon' interactions are measured.
$<n_{V^{\circ}}>_A$ =0.095$\pm$0.011 and $<n_{V^{\circ}}>_N$ =
0.059$\pm$0.012, respectively. The multiplicity gain of
$V^{\circ}$ particles in nuclear interactions, measured as the
ratio of $\rho_{V^{\circ}} = <n_{V^{\circ}}>_A/<n_{V^{\circ}}>_N =
1.61 \pm 0.23$, compared to that for ${\pi}^-$ mesons,
${\rho}_{{\pi}^-} = 1.10 \pm 0.03$, indicates that the nuclear
influence on the strange particle production is larger than
for non-strange one.\\ 
The measured value
of the difference ${\delta}_{V^{\circ}}^{exp} = <n_{V^{\circ}}>_A
- <n_{V^{\circ}}>_N = 0.037 \pm 0.010$ is mainly contributed by
the multiplicity gain in the backward hemisphere ($x_F < 0$)
and can be approximately described by
a simple model incorporating the secondary
inelastic interactions of produced pions within the nucleus.

\noindent {\bf{Acknowledgement:}} The activity of one of the
authors (Zh.K.) is supported by Cooperation Agreement between DESY
and YerPhI signed on December 6, 2002.


\end{document}